\documentclass[prd, preprint, longbibliography, 12pt]{revtex4-1}

\usepackage{commath}
\usepackage{amsmath}
\usepackage{amssymb}
\usepackage{setspace}
\usepackage{graphicx}
\usepackage{natbib}
\usepackage{float}
\usepackage{tensor}
\usepackage[utf8]{inputenc}
\usepackage{amsfonts}
\usepackage{braket}
\usepackage{esint}
\usepackage{breqn}
\usepackage{IEEEtrantools}
\usepackage{hyperref}

\usepackage[english]{babel}
\usepackage{graphicx}
\usepackage{amsmath}

\usepackage{amssymb}
\usepackage{multirow}
\usepackage{url}
\usepackage{tikz}
\usepackage{braket}
\usetikzlibrary{decorations.markings}
\usepackage{subcaption}
\usepackage{cancel}
\usepackage{mathtools}
\usepackage[utf8]{inputenc}
\usepackage{xargs, ifthen}
\usepackage[breakwords]{truncate}

\usepackage{hyperref}
\DeclareMathAlphabet{\pazocal}{OMS}{zplm}{m}{n}

\allowdisplaybreaks

\def\s[#1\s]{\begin{align}\begin{split}#1\end{split}\end{align}}
\def\[#1\]{\begin{align}#1\end{align}}

\begin{document}
 
 %

\begin{center}
{ \large \bf Why does the Kerr-Newman black hole have the same gyromagnetic ratio as the electron?}



\vskip 0.2 in

{\large{\bf Meghraj M S$^{a1}$, Abhishek Pandey$^{b2}$ and Tejinder P.  Singh$^{c3}$ }}

\medskip

{\it $^a$Indian Institute of Science Education and Research Kolkata, 741246, India}\\

{\it $^b$Birla Institute of Technology and Science, Pilani, 333031, India}\\

{\it $^c$Tata Institute of Fundamental Research, Homi Bhabha Road, Mumbai 400005, India}\\

\bigskip

 \; {$^{1}$\tt mms15ms051@iiserkol.ac.in}, {$^{2}$\tt f2016588@pilani.bits-pilani.ac.in}, \; {$^3$\tt tpsingh@tifr.res.in}

\end{center}

\bigskip

\centerline{\bf ABSTRACT}
\noindent  We have recently proposed a deterministic matrix dynamics at the Planck scale, for gravity coupled to Dirac fermions, evolving in the so-called Connes time. By coarse-graining this dynamics over time intervals much larger than Planck time, we derived the space-time manifold,  quantum theory,  and classical general relativity, as low energy emergent approximations to the underlying matrix dynamics. In the present article, we show how to include Yang-Mills gauge fields in this Planck scale matrix dynamics. We do this by appropriately modifying the fundamental action for the previously introduced `atom' of space-time-matter [which we now call an `aikyon']. This is achieved by modifying the Dirac operator to include a `potential' for the Yang-Mills aspect, and a `current' for the Yang-Mills charge. Our work opens up an avenue for unification of gravity with gauge-fields and Dirac fermions. We show how spontaneous localisation in the matrix dynamics  gives rise to general relativity coupled to 
gauge-fields and relativistic point particles, in the classical limit. We use this formalism to explain the remarkable fact that the Kerr-Newman black hole has the same value for the gyromagnetic ratio as that for a Dirac fermion, both being twice the classical value.

\bigskip

\bigskip

\section{Introduction}
We have recently proposed a deterministic matrix dynamics at the Planck scale \cite{maithresh2019}. The motivation for having such a dynamics is to find an equivalent reformulation of quantum (field) theory which does not refer to classical time. To arrive at such a dynamics we start from a Riemannian space-time manifold endowed with a metric and a connection, and having as matter source relativistic point particles. To move towards the said matrix dynamics, we raise space-time points to the status of operators [equivalently matrices]. The Dirac operator on the manifold is employed to construct a gravity operator, and the c-numbers which describe material point particles are also raised to the status of operators. However, these operators do not obey quantum commutation relations nor the laws of quantum (field) theory. Instead, they obey a Lagrangian matrix dynamics, assumed to operate at the Planck scale. In this dynamics, there is no space-time; in fact there is no distinction between matter and space-time any longer. An elementary particle and its gravitation together constitute an `atom' of space-time-matter, which we shall henceforth call an `aikyon' [derived from the Sanskrit word `aikya', meaning oneness]. Thus an electron-aikyon is an electron together with the gravitation it produces, with no reference to any background space-time in which the electron might appear to be embedded. Every aikyon has an associated length scale, but no mass nor spin, these being emergent concepts. We constructed an action principle for an aikyon, and the total action for many aikyons is the sum of their individual actions. The Lagrange equations of motion for the aikyons, as well as the equivalent Hamilton equations of motion, can be derived from extremising the action. These equations of motion describe evolution in Connes time, which is a reversible time parameter present in the matrix dynamics, because of a global unitary invariance of the dynamics [and its connection with non-commutative geometry]. The eigenvalues of the Dirac operator capture information about the metric and the induced curvature, in this non-commutative matrix dynamics \cite{Rovelli, Landi1999}.

An aikyon is described by a matrix made of complex-valued Grassmann numbers, and denoted as $q$. Any Grassmann matrix can be written as a sum of a `bosonic' matrix $q_B$ made of even grade elements of the Grassmann algebra, and a `fermionic' matrix $q_F$ made of odd grade elements of the Grassmann algebra. Thus $q\equiv q_B + q_F$, implying that an aikyon is neither a boson nor a fermion, but should be a thought of as a combination of the two, with the bosonic part describing the gravitation aspect, and the fermionic part representing the matter aspect. We have proposed the following action principle for an aikyon:
\begin{equation}
 \frac{S}{C_0}  =  \frac{1}{2} \int \frac{d\tau}{\tau_{Pl}} \; Tr \bigg[\frac{L_P^2}{L^2c^2}\; \left(\dot{q}_B +\beta_1 \frac{L_P^2}{L^2}\dot{q}_F\right)\; \left(\dot{q}_B +\beta_2 \frac{L_P^2}{L^2} \dot{q}_F\right) \bigg]
 \label{acnaik}
\end{equation}
where $\beta_1$ and $\beta_2$ are  constant self-adjoint fermionic matrices. These matrices make the Lagrangian bosonic.
The only two fundamental constants are Planck length and Planck time - these scale the length scale $L$ of the aikyon, and the Connes time, respectively. $C_0$ is a constant with dimensions of action, which will be identified with Planck's constant in the emergent theory. The Lagrangian and action are not restricted to be self-adjoint. A dot denotes derivative with respect to Connes time. By varying this action w.r.t. $q_B$ and $q_F$ one gets a pair of coupled equations of motion, which can be solved to find the evolution of $q_B$ and $q_F$. The respective momenta $p_B$ and $p_F$ are constants of motion, and the expression for $p_B$ can be written as an eigenvalue equation for the modified Dirac operator $D\equiv D_B + D_F$:
\begin{equation}
\left[D_B + D_F\right] \psi =  \frac{1}{L} \bigg(1+ i \frac{L_P^2}{L^2}\bigg)\psi
\label{dfull}
\end{equation}
where
\begin{equation}
D_B \equiv \frac{1}{Lc}\;  \frac{dq_B}{d\tau}; \qquad   D_F \equiv \frac{L_P^2}{L^2}\frac{\beta_1+\beta_2}{2Lc}\;  \frac{dq_F}{d\tau}
\end{equation}
$D_B$ is defined such that in the commutative c-number limit where space-time emerges, it becomes the standard Dirac operator on a Riemannian manifold. $D_F$ is defined such that upon spontaneous localisation,  it gives rise to the classical action for a relativistic point particle.

The Hilbert space in which this matrix dynamics operates at the Planck scale is populated by a large number of aikyons, labelled $q_i$, whose total action can be symbolically written in terms of their respective modified Dirac operators $D_i$:
\begin{equation}
\frac{S_{total}}{C_0} = \frac{1}{2} \int \frac{d\tau}{\tau_{Pl}}\; \sum_i \ Tr\; [D_i^2]
\end{equation}
The generalised Dirac operator $D$ is not self-adjoint, because $D_F$ is not self-adjoint, even though $D_B$ is. Related to this is the fact that the Hamiltonian of the theory is in general not self-adjoint either. The anti-self-adjoint part is however negligible if the length scale $L$ for every aikyon is much larger than Planck length, and the number $N_c$ of aikyons which are entangled with each other are much smaller than a certain critical limit, which is of the order $N_c \sim L/L_P$.

Next, we ask what does the emergent dynamics look like, if we are not observing this matrix dynamics at Planck time resolution, but coarse-grained over time intervals much larger than Planck time? This question can be answered by employing the methods of statistical thermodynamics, and assuming that an ensemble of microstates describes various possible motions at Planck scale resolution, all of which give rise, at equilibrium, to the same macrostate \cite{Adler:04, Adler:94, Adler-Millard:1996}. The macrostate is determined by the conventional methods of statistical mechanics, by maximising the combinatorial entropy constructed from a probability distribution in the phase space for this matrix dynamics. The physics is the same as when we find the emergent thermodynamic state from the statistical mechanics of a molecular fluid, when we are not examining the fluid at the level of its molecular resolution, but at a 
coarse-grained level obtained by averaging over length scales  much larger than inter-molecular separation. 

The emergent dynamics falls into two limiting classes, depending on the degree of entanglement between different aikyons. The first limiting class is obtained when the $L$ values for all the aikyons are much larger than Planck length, and the number of entangled aikyons in the system is much smaller than the critical number. 
When this happens, the anti-self-adjoint part of the full Hamiltonian is negligible.
In this limit, one recovers the sought for space-time free limit of quantum theory. Evolution is still described in Connes time. But the (averaged) canonical variables now obey quantum commutation relations, separately for the bosonic and fermionic parts, and Planck's constant $\hbar$ emerges. One also recovers Heisenberg equations of motion for the dynamical degrees of freedom, again separately for bosons and fermions. There is an equivalent Schr\"{o}dinger picture. This emergent dynamics is a quantum gravity, and comes into play whenever we want to find out the gravitational effect of a quantum system non-perturbatively, at energies below Planck scale. For instance, if we were to ask for the gravitational effect of an electron in the double-slit experiment.

The opposite  emergent limit is when a sufficiently large number of aikyons get entangled ($N\sim N_c$): under these circumstances an effective length scale $L_{eff} \sim L/N$  goes below Planck length, and the anti-self-adjoint part of the Hamiltonian becomes significant. The entangled system of aikyons undergoes spontaneous localisation to a classical state. This leads to the emergence of the space-time manifold, spatial localisation of macroscopic objects, and the laws of classical general relativity. The total action for the matrix dynamics of the entangled aikyons is reduced to the action for classical general relativity, as a consequence of spontaneous localisation:
\begin{equation}
S = \int d\tau \sum_i Tr D^2_i  \quad  {\mathbf \longrightarrow} \quad \int d\tau  \int d^4x\; \sqrt{g} \; \bigg [\frac{c^3}{2G}R + c\; \sum_i m_i \delta^3({\bf x} - {\bf x_0})\bigg]
\label{pg}
\end{equation}
In deriving this result, key use is made of a theorem in Riemannian geometry, which relates the trace of the squared Dirac operator $D_B^2$ to the Einstein-Hilbert action:
\begin{equation}
Tr\ [L_P^2\ D_B^2] \propto L_P^{-2}\int d^4x\ \sqrt{g} \ R \quad + {\cal O}(L_P^{0})
\end{equation}
This relation comes about from the so-called heat kernel series expansion of the left hand side,  in powers of $L_P^{-2}$
\cite{Connes2000}.

Given a classical space-time dominated by classical material bodies [as opposed to being dominated by quantum objects], one can reformulate the first limiting class above [quantum theory without classical time] as quantum (field) theory on this background space-time manifold. One sees in this way how quantum theory and classical general relativity emerge as low energy approximations, being two opposite limits of the emergent dynamics [low entanglement limit, and high entanglement limit, respectively]. 

Of course a limitation of the above dynamics is that it considers only gravity and Dirac fermions (albeit, unified in the aikyon concept in the Planck scale matrix dynamics). In the present article we show to include Yang-Mills gauge fields in our approach. 
For achieving this goal, we are guided by a few principles. Firstly, we would not like to lose out on the aikyon concept, which unifies a particle with its gravitation, by expressing them as $q=q_B + q_F$. As we have seen, what appears in the above action are not $q_B$ and $q_F$ themselves, but their time derivatives. These time derivatives are respectively identified with gravity and with the source of gravity, namely the material particles. The structure is symbolically of the form:
\begin{equation}
D^2 \equiv [D_B + D_F]^2 \sim D_B^2 + D_B\; D_F + D_F^2
\end{equation}
The first term on the right, i.e. $D_B^2$, leads to the gravity part of the Einstein-Hilbert action, whereas the second term, i.e. the cross-term $D_B\; D_F$, gives the relativistic point particle as the matter source. The last term, $D_F^2$, is a higher order term in the $L_P^2$ expansion, which we ignore for the time-being. Einstein equations have the remarkable property that they are linear in the source mass, despite being second order equations. This makes them unlike the Klein-Gordon equation which is second order, and also quadratic in the source mass. In this respect Einstein's equations are closer to the Dirac equation which is linear in the source mass. This linear dependence on the mass in Einstein equations is easily understood in our theory, because they originate from the first order equation (\ref{dfull}) and the mass term arises from the term linear in $D_F$ in the above expansion of $D^2$. 

We also know that when spontaneous localisation localises the fermionic part of an aikyon to a specific position, the associated space-time manifold, metric, and gravitational field, emerge concurrently with the localisation. The same feature will have to be true for the Yang-Mills field produced by its associated charge and its current: the localised current must emerge concurrently with its associated gauge field, and these two must emerge concurrently with the localised mass and associated gravitation of the aikyon.

These remarks guide us as to how Yang-Mills fields can be included. We know that in the Dirac equation they enter as an `internal' connection in the form: $D_B \longrightarrow D_B + \alpha A$. Since in our matrix dynamics $D_B$ is identified with the velocity $dq_B/d\tau$, we propose to identify the (self-adjoint) gauge-field potential operator $A$ with $q_B$. This way we make the gauge-field and gravitation respectively the position and velocity aspects of the aikyon. Similarly, since $D_F$, which gives rise to the source mass term, is identified with the velocity $dq_F/d\tau$, we propose to identify the current $j$  (which is the source of the gauge field) with $q_F$. Thus, the new squared Dirac operator will be of the form
\begin{equation}
D_{new}^2 \equiv [D_B + \alpha A + D_F + j]^2 \sim D_B^2 + \alpha^2 A^2 + \alpha  D_B A + D_B D_F + D_B j + \alpha A D_F + \alpha Aj + D_F^2 + j^2 +D_F j
\end{equation}
This includes the action term  for the gauge fields, symbolically written as $\alpha^2 A^2$, and their source term $D_{B} j$ (which is linear in the current) apart from the gravitation terms we already have earlier, and a few new terms. Hence, by including the gauge-fields and their charges in the action for the aikyon, we will derive Einstein equations with gauge fields and material particles as source. We will also derive quantum field theory for these gauge fields coupled to Dirac fermions. 

In arriving at Einstein equations coupled to gauge-fields, after spontaneous localisation  from this matrix dynamics, we will make use of the following result from geometry, for the heat kernel expansion of the bosonic part $D_{Bnew} \equiv D_B + \alpha A$ of $D_{new}$:
\begin{equation}
Tr\ [L_P^2\ D_{Bnew}^2 ]\propto L_P^{-2}\int d^{4}x\sqrt{g}\left({R + L_{P}^{2} \, \alpha^{2} F^{i}_{\mu\nu} F_{i}^{\mu\nu}}\right)
+ {\cal O}(L_P^{2})
\end{equation}
This generalises the corresponding result above, when there is only gravity, represented by $D_B^2$, but no gauge-field $A$.
In this context, we quote from the work of Chamseddine and Connes \cite{Chams:1997} who discuss the spectral action when Yang-Mills fields are included:

"
It is, also possible to introduce a mass scale $m_{0}$ and consider $\chi$ to be a function of the dimensionless variable $\chi\left(\dfrac{P}{m_{0}^{2}}\right)$. In this case terms coming from $a_n(P)$, $n>4$ will be suppressed by the powers of $\dfrac{1}{m_{0}^{2}}$ :
\s[
I_{b} &= \dfrac{N}{48 \pi^{2}} \biggl[\biggr. 12 m_{0}^{4} f_{0} \int d^{4}x\sqrt{g} + m_{0}^{2} f_{2} \int d^{4}x\sqrt{g} R\\
&+ f_{4} \int d^{4}x\sqrt{g} \biggl[\biggr. -\dfrac{3}{20} C_{\mu \nu \rho \sigma} C^{\mu \nu \rho \sigma} + \dfrac{11}{20} R^{*} R^{*} + \dfrac{1}{10} R;_{\mu}\, ^{\mu}\\
&+ \dfrac{g^{2}}{N} F^{i}_{\mu\nu} F^{\mu\nu i} \biggl.\biggr] + \mathcal{O}\biggl(\biggr. \dfrac{1}{m_{0}^{2}} \biggl.\biggr) \biggl.\biggr]
\s]
where
\begin{itemize}
\item $\dfrac{N m_{0}^{2} f_{2}}{48 \pi^{2}} \int d^{4}x\sqrt{g} R$ term is the Einstein-Hilbert action
\item $\dfrac{N m_{0}^{4} f_{0}}{4 \pi^{2}} \int d^{4}x\sqrt{g}$ term is responsible for the cosmological constant
\item $\dfrac{f_{4} g^{2}}{48 \pi^{^{2}}} \int d^{4}x\sqrt{g} F^{i}_{\mu\nu} F^{\mu\nu i}$ term is the Yang-Mills action
\item $-\dfrac{N f_{4}}{320 \pi^{^{2}}} \int d^{4}x\sqrt{g} C_{\mu \nu \rho \sigma} C^{\mu \nu \rho \sigma}$ term would be responsible for the Conformal gravity
\item $\dfrac{11 N f_{4}}{960 \pi^{^{2}}} \int d^{4}x\sqrt{g} R^{*} R^{*}$ term would be responsible for the Gauss-Bonnet gravity"
\end{itemize}
This is the expansion of the squared Dirac operator when gauge fields are included alongside gravity. In our case, we set the scale $m_0$ to be the inverse of Planck length. Also, we do not take into account the volume term, and  conformal gravity, and Gauss-Bonnet gravity in our present work. Note though that their analysis is classical; whereas we will employ it to construct a matrix dynamics from which quantum theory emerges.

When spontaneous localisation results in the localisation of the fermionic part of an aikyon, it simultaneously gives rise to the space-time manifold as well as the gravitational field. Now, if the fermion has a charge, such as an electric charge, the associated electromagnetic field must also appear along with gravitation. That is the goal accomplished in the present work. From this point of view, it is difficult to come to the conclusion that the space-time manifold and its associated curvature are in any sense more fundamental than the gauge fields which supposedly `live' `on' space-time. The gauge-fields of a fermion, and their associated charge $\alpha$,  are as fundamental as space-time-gravitation of the fermion, with its associated length $L$.
No more, no less.

\section{Euler-Lagrange equations for the  Trace Lagrangian including Yang-Mills fields}
\label{sec:EL}
We generalise the previous action (\ref{acnaik}) of an aikyon,  to $S=\int d\tau \; {\cal L}$, where the new Lagrangian is given by
\s[
\mathcal{L} = Tr \biggl[\biggr. \dfrac{L_{p}^{2}}{L^{4}} \biggl\{\biggr. i\alpha \biggl(\biggr. q_{B} + \dfrac{L_{p}^{2}}{L^{2}}\beta_{1} q_{F} \biggl.\biggr) + L \biggl(\biggr. \dot{q}_{B} + \dfrac{L_{p}^{2}}{L^{2}}\beta_{1}\dot{q}_{F} \biggl.\biggr) \biggl.\biggr\}\\
\biggl\{\biggr. i\alpha \biggl(\biggr. q_{B} + \dfrac{L_{p}^{2}}{L^{2}}\beta_{2} q_{F} \biggl.\biggr) + L \biggl(\biggr. \dot{q}_{B} + \dfrac{L_{p}^{2}}{L^{2}}\beta_{2} \dot{q}_{F} \biggl.\biggr) \biggl.\biggr\} \biggl.\biggr]
\s]
This Lagrangian for an aikyon should be compared with the earlier one which had only gravity and Dirac fermions as unified components of the aikyon. This new Lagrangian here also includes gauge-fields and their currents, through $q_B$ and $q_F$,  as we will justify further, subsequently. $\alpha$ is the Yang-Mills coupling constant, here assumed to be a real number. It appears to us that a more general treatment would have $\alpha$ as a matrix - we leave this consideration for future work.  Gravitation, and Yang-Mills fields, and their corresponding sources,  are unified here as the `position' $q$ and `velocity' $dq/d\tau$ of the aikyon. With position being the Yang-Mills part, and velocity being the gravitation part. Naively, it might appear that position should go with gravitation, and velocity with Yang-Mills. On the other hand, the Dirac operator $D_B$ is conventionally thought of as momentum, and is also related to gravitation; so it seems reasonable to associate gravitation with velocity. A closer look reveals this to be akin to the Lagrangian for a harmonic oscillator, as is reflected also in the solutions we find below. This quadratic form also suggests that inclusion of higher order terms in the heat-kernel expansion, beyond order $L_P^0$, will reveal departures from harmonic oscillator behaviour, and equations of motion that are higher than second order.

A word about dimensions. $q$ has dimensions of length. The time derivative denoted by the dot stands for $c\; d\tau$, so that the velocity is dimensionless. The coupling constant $\alpha$ is dimensionless, and so are the Lagrangian and the action. Also, the Dirac operator $D_B$ is defined as before: $D_B = (1/L)\dot{q}_B$, and the relation between the gauge potential $A$ and 
$q_B$ is $ A L^2 \equiv \alpha q_B$. Hence, $D_{Bnew} = D_B + \alpha  q_B/L^2$ which is a self-adjoint operator.

On expanding this Lagrangian, we obtain sixteen terms, which are as follows:
 \s[
\mathcal{L} = Tr \biggl[\biggr. \dfrac{L_{p}^{2}}{L^{4}} \biggl\{\biggr. -\alpha^{2} \; q_{B}q_{B} - \alpha^{2} \dfrac{L_{p}^{2}}{L^{2}} \; q_{B}\beta_{2}q_{F} + i\alpha L \; q_{B}\dot{q}_{B} + i\alpha \dfrac{L_{p}^{2}}{L} \; q_{B}\beta_{2} \dot{q}_{F}\\
-\alpha^{2} \dfrac{L_{p}^{2}}{L^{2}} \; \beta_{1}q_{F}q_{B} - \alpha^{2} \dfrac{L_{p}^{4}}{L^{4}} \; \beta_{1}q_{F}\beta_{2}q_{F} + i\alpha \dfrac{L_{p}^{2}}{L} \; \beta_{1}q_{F}\dot{q}_{B} + i\alpha \dfrac{L_{p}^{4}}{L^{3}} \; \beta_{1}q_{F}\beta_{2}\dot{q}_{F}\\
+i\alpha L \; \dot{q}_{B}q_{B} + i\alpha \dfrac{L_{p}^{2}}{L} \; \dot{q}_{B}\beta_{2}q_{F} + L^{2} \; \dot{q}_{B}\dot{q}_{B} + L_{P}^{2} \; \dot{q}_{B}\beta_{2}\dot{q}_{F}\\
+i\alpha \dfrac{L_{p}^{2}}{L} \; \beta_{1}\dot{q}_{F}q_{B} + i\alpha \dfrac{L_{p}^{4}}{L^{3}} \; \beta_{1}\dot{q}_{F}\beta_{2}q_{F} + L_{P}^{2} \; \beta_{1}\dot{q}_{F}\dot{q}_{B} + \dfrac{L_{p}^{4}}{L^{2}} \; \beta_{1}\dot{q}_{F}\beta_{2}\dot{q}_{F} \biggl.\biggr\} \biggl.\biggr]
\s]
Now in order to write down the Euler-Lagrange equation for $q_{B}$, we need $\dfrac{\partial \mathcal{L}}{\partial \dot{q}_{B}}$ and $\dfrac{\partial \mathcal{L}}{\partial q_{B}}$, the calculations for which are as shown below. The trace derivative is employed for carrying out differentiation with respect to a matrix, as in the theory of trace dynamics \cite{Adler:04}.
\s[
\dfrac{\partial \mathcal{L}}{\partial \dot{q}_{B}} = i\alpha L \, q_{B} + i\alpha \dfrac{L_{p}^{2}}{L} \, \beta_{1} q_{F} + L^{2}\, \dot{q}_{B} + L_{P}^{2} \, \beta_{1}\dot{q}_{F}\\
+i\alpha L \, q_{B} + i\alpha \dfrac{L_{p}^{2}}{L} \, \beta_{2} q_{F} + L^{2}\, \dot{q}_{B} + L_{P}^{2} \, \beta_{2}\dot{q}_{F}
\s]
\[
\implies \dfrac{\partial \mathcal{L}}{\partial \dot{q}_{B}} = 2(i\alpha L \, q_{B} + L^{2}\, \dot{q}_{B}) + \dfrac{L_{p}^{2}}{L^{2}}(\beta_{1}+\beta_{2}) \, [i\alpha L \, q_{F} + L^{2} \, \dot{q}_{F}]
\]
\s[
\dfrac{\partial \mathcal{L}}{\partial q_{B}} = -\alpha^{2} \, q_{B} - \alpha^{2} \dfrac{L_{p}^{2}}{L^{2}} \, \beta_{1}q_{F} + i\alpha L \, \dot{q}_{B} + i\alpha \dfrac{L_{p}^{2}}{L} \, \beta_{1}\dot{q}_{F}\\
-\alpha^{2} \, q_{B} - \alpha^{2} \dfrac{L_{p}^{2}}{L^{2}} \, \beta_{2}q_{F} + i\alpha L \, \dot{q}_{B} + i\alpha \dfrac{L_{p}^{2}}{L} \, \beta_{2}\dot{q}_{F}
\s]
\[
\implies \dfrac{\partial \mathcal{L}}{\partial q_{B}} = 2(-\alpha^{2} \, q_{B} + i\alpha L \, \dot{q}_{B}) + \dfrac{L_{p}^{2}}{L^{2}}(\beta_{1}+\beta_{2}) \, [-\alpha^{2} \, q_{F} + i\alpha L \, \dot{q}_{F}]
\]
Now taking derivative of $\dfrac{\partial \mathcal{L}}{\partial \dot{q}_{B}}$ with respect to Connes time, we obtain:
\[
\dfrac{d}{d\tau} \biggl(\biggr. \dfrac{\partial \mathcal{L}}{\partial \dot{q}_{B}} \biggl.\biggr) = 2(i\alpha L \, \dot{q}_{B} + L^{2} \, \ddot{q}_{B}) + \dfrac{L_{p}^{2}}{L^{2}}(\beta_{1}+\beta_{2}) \, [i\alpha L \, \dot{q}_{F} + L^{2} \, \ddot{q}_{F}]
\]
This tells us that the Euler-Lagrange equation for $q_{B}$ is as follows:
\[
\dfrac{d}{d\tau} \biggl(\biggr. \dfrac{\partial \mathcal{L}}{\partial \dot{q}_{B}} \biggl. \biggr) = \dfrac{\partial \mathcal{L}}{\partial q_{B}}
\]
\s[
\implies 2(\cancel{i\alpha L \, \dot{q}_{B}} + L^{2} \, \ddot{q}_{B}) + \dfrac{L_{p}^{2}}{L^{2}}(\beta_{1}+\beta_{2}) \, [\cancel{i\alpha L \, \dot{q}_{F} }+ L^{2} \, \ddot{q}_{F}] =\\
2(-\alpha^{2} \, q_{B} + \cancel{i\alpha L \, \dot{q}_{B}}) + \dfrac{L_{p}^{2}}{L^{2}}(\beta_{1}+\beta_{2}) \, [-\alpha^{2} \, q_{F} + \cancel{i\alpha L \, \dot{q}_{F}}]
\s]
This tells us that the Euler-Lagrange equation for $q_{B}$ is
\[
\boxed{
\ddot{q}_{B} + \dfrac{\alpha^{2}}{L^{2}}q_{B} = -\dfrac{L_{p}^{2}}{L^{2}} \biggl(\biggr.\dfrac{\beta_{1}+\beta_{2}}{2}\biggl.\biggr) \biggl[\biggr.\ddot{q}_{F} + \dfrac{\alpha^{2}}{L^{2}}q_{F} \biggl.\biggr]}
\label{eq:eleqn1}
\]
Now in order to write down the Euler-Lagrange equation for $q_{F}$, we need $\dfrac{\partial \mathcal{L}}{\partial \dot{q}_{F}}$ and $\dfrac{\partial \mathcal{L}}{\partial q_{F}}$ the calculations for which are as shown below:
\s[
\dfrac{\partial \mathcal{L}}{\partial \dot{q}_{F}} = i\alpha \dfrac{L_{p}^{2}}{L} \, q_{B}\beta_{2} + i\alpha \dfrac{L_{p}^{4}}{L^{3}} \, \beta_{1}q_{F}\beta_{2} + L_{P}^{2} \, \dot{q}_{B}\beta_{2} + \dfrac{L_{p}^{4}}{L^{2}} \, \beta_{1}\dot{q}_{F}\beta_{2}\\
+i\alpha \dfrac{L_{p}^{2}}{L} \, q_{B}\beta_{1} + i\alpha \dfrac{L_{p}^{4}}{L^{3}} \, \beta_{2}q_{F}\beta_{1} + L_{P}^{2} \, \dot{q}_{B}\beta_{1} + \dfrac{L_{p}^{4}}{L^{2}} \, \beta_{2}\dot{q}_{F}\beta_{1}
\s]
\[
\implies \dfrac{\partial \mathcal{L}}{\partial \dot{q}_{F}} = L_{P}^{2} \biggl[\biggr. \dfrac{i\alpha}{L} \, q_{B} + \dot{q}_{B} \biggl.\biggr] (\beta_{1}+ \beta_{2}) + i\alpha \dfrac{L_{p}^{4}}{L^{3}} \, (\beta_{1} q_{F}\beta_{2} + \beta_{2} q_{F}\beta_{1}) + \dfrac{L_{p}^{4}}{L^{2}} \, (\beta_{1}\dot{q}_{F}\beta_{2} + \beta_{2}\dot{q}_{F}\beta_{1})
\]
\s[
\dfrac{\partial \mathcal{L}}{\partial q_{F}} = -\alpha^{2}\dfrac{L_{p}^{2}}{L^{2}} \, q_{B}\beta_{2} - \alpha^{2} \dfrac{L_{p}^{4}}{L^{4}} \, \beta_{1}q_{F}\beta_{2} + i\alpha \dfrac{L_{p}^{2}}{L} \, \dot{q}_{B}\beta_{2} + i\alpha \dfrac{L_{p}^{4}}{L^{3}} \, \beta_{1}\dot{q}_{F}\beta_{2}\\
-\alpha^{2}\dfrac{L_{p}^{2}}{L^{2}} \, q_{B}\beta_{1} - \alpha^{2} \dfrac{L_{p}^{4}}{L^{4}} \, \beta_{2}q_{F}\beta_{1} + i\alpha \dfrac{L_{p}^{2}}{L} \, \dot{q}_{B}\beta_{1} + i\alpha \dfrac{L_{p}^{4}}{L^{3}} \, \beta_{2}\dot{q}_{F}\beta_{1}
\s]
\[
\implies \dfrac{\partial \mathcal{L}}{\partial q_{F}} = L_{P}^{2}\biggl[\biggr. -\dfrac{\alpha^{2}}{L^{2}} \, q_{B} + \dfrac{i\alpha}{L} \, \dot{q}_{B} \biggl.\biggr] (\beta_{1}+ \beta_{2}) + i\alpha \dfrac{L_{p}^{4}}{L^{3}} \, (\beta_{1}\dot{q}_{F}\beta_{2} + \beta_{2}\dot{q}_{F}\beta_{1}) - \alpha^{2} \dfrac{L_{p}^{4}}{L^{4}} \, (\beta_{1}q_{F}\beta_{2} + \beta_{2}q_{F}\beta_{1})
\]
Now taking derivative of $\dfrac{\partial \mathcal{L}}{\partial \dot{q}_{F}}$ with respect to Connes' time, we obtain:
\[
\dfrac{d}{d\tau} \biggl(\biggr. \dfrac{\partial \mathcal{L}}{\partial \dot{q}_{F}} \biggl.\biggr) = L_{P}^{2} \biggl[\biggr. \dfrac{i\alpha}{L} \, \dot{q}_{B} + \ddot{q}_{B} \biggl.\biggr] (\beta_{1}+ \beta_{2}) + i\alpha \dfrac{L_{p}^{4}}{L^{3}} \, (\beta_{1}\dot{q}_{F}\beta_{2} + \beta_{2}\dot{q}_{F}\beta_{1}) + \dfrac{L_{p}^{4}}{L^{2}} \, (\beta_{1}\ddot{q}_{F}\beta_{2} + \beta_{2}\ddot{q}_{F}\beta_{1})
\]
This tells us that the Euler-Lagrange equation for $q_{F}$ is as follows:
\[
\dfrac{d}{d\tau} \biggl(\biggr. \dfrac{\partial \mathcal{L}}{\partial \dot{q}_{F}} \biggl.\biggr) = \dfrac{\partial \mathcal{L}}{\partial q_{F}}
\]
\s[
\implies L_{P}^{2} \biggl[\biggr. \cancel{\dfrac{i\alpha}{L} \, \dot{q}_{B}} + \ddot{q}_{B} \biggl.\biggr] (\beta_{1}+ \beta_{2}) + \cancel{i\alpha \dfrac{L_{p}^{4}}{L^{3}} \, (\beta_{1}\dot{q}_{F}\beta_{2} + \beta_{2}\dot{q}_{F}\beta_{1})} + \dfrac{L_{p}^{4}}{L^{2}} \, (\beta_{1}\ddot{q}_{F}\beta_{2} + \beta_{2}\ddot{q}_{F}\beta_{1}) =\\
L_{P}^{2}\biggl[\biggr. -\dfrac{\alpha^{2}}{L^{2}} \, q_{B} + \cancel{\dfrac{i\alpha}{L} \, \dot{q}_{B}} \biggl.\biggr] (\beta_{1}+ \beta_{2}) + \cancel{i\alpha \dfrac{L_{p}^{4}}{L^{3}} \, (\beta_{1}\dot{q}_{F}\beta_{2} + \beta_{2}\dot{q}_{F}\beta_{1})} - \alpha^{2} \dfrac{L_{p}^{4}}{L^{4}} \, (\beta_{1}q_{F}\beta_{2} + \beta_{2}q_{F}\beta_{1})
\s]
This tells us that the Euler-Lagrange equation for $q_{F}$ is
\[
\boxed{
\biggl[\biggr. \ddot{q}_{B} + \dfrac{\alpha^{2}}{L^{2}}q_{B} \biggl.\biggr] (\beta_{1}+\beta_{2}) = -\dfrac{L_{P}^{2}}{L^{2}} \biggl(\biggr. \beta_{1} \biggl[\biggr. \ddot{q}_{F} + \dfrac{\alpha^{2}}{L^{2}}q_{F} \biggl.\biggr] \beta_{2} +  \beta_{2} \biggl[\biggr. \ddot{q}_{F} + \dfrac{\alpha^{2}}{L^{2}}q_{F} \biggl.\biggr] \beta_{1} \biggl.\biggr)}
\label{eq:eleqn2}
\]
Now trying to solve the two Euler-Lagrange equations (\ref{eq:eleqn1}) and (\ref{eq:eleqn2}) by substituting one in the other we obtain:
\[
-\dfrac{(\beta_{1}+\beta_{2})}{2} \biggl[\biggr. \ddot{q}_{F} + \dfrac{\alpha^{2}}{L^{2}}q_{F} \biggl.\biggr] (\beta_{1}+\beta_{2}) + \beta_{1} \biggl[\biggr. \ddot{q}_{F} + \dfrac{\alpha^{2}}{L^{2}}q_{F} \biggl.\biggr] \beta_{2} +  \beta_{2} \biggl[\biggr. \ddot{q}_{F} + \dfrac{\alpha^{2}}{L^{2}}q_{F} \biggl.\biggr] \beta_{1} = 0
\]
\s[
\implies -\dfrac{1}{2} \beta_{1} \ddot{q}_{F} \beta_{1} - \dfrac{\alpha^{2}}{2 L^{2}} \beta_{1} q_{F} \beta_{1} - \dfrac{1}{2} \beta_{1} \ddot{q}_{F} \beta_{2} - \dfrac{\alpha^{2}}{2 L^{2}} \beta_{1} q_{F} \beta_{2}\\
- \dfrac{1}{2} \beta_{2} \ddot{q}_{F} \beta_{1} - \dfrac{\alpha^{2}}{2 L^{2}} \beta_{2} q_{F} \beta_{1} - \dfrac{1}{2} \beta_{2} \ddot{q}_{F} \beta_{2} - \dfrac{\alpha^{2}}{2 L^{2}} \beta_{2} q_{F} \beta_{2}\\
+ \beta_{1} \ddot{q}_{F} \beta_{2} + \dfrac{\alpha^{2}}{L^{2}} \beta_{1} q_{F} \beta_{2} + \beta_{2} \ddot{q}_{F} \beta_{1} + \dfrac{\alpha^{2}}{L^{2}} \beta_{2} q_{F} \beta_{1} = 0
\s]
\s[
\implies -\dfrac{1}{2} \beta_{1} \ddot{q}_{F} \beta_{1} - \dfrac{\alpha^{2}}{2 L^{2}} \beta_{1} q_{F} \beta_{1} + \dfrac{1}{2} \beta_{1} \ddot{q}_{F} \beta_{2} + \dfrac{\alpha^{2}}{2 L^{2}} \beta_{1} q_{F} \beta_{2}\\
+ \dfrac{1}{2} \beta_{2} \ddot{q}_{F} \beta_{1} + \dfrac{\alpha^{2}}{2 L^{2}} \beta_{2} q_{F} \beta_{1} - \dfrac{1}{2} \beta_{2} \ddot{q}_{F} \beta_{2} - \dfrac{\alpha^{2}}{2 L^{2}} \beta_{2} q_{F} \beta_{2} = 0
\s]
\[
\implies -\dfrac{(\beta_{1}-\beta_{2})}{2} \biggl[\biggr. \ddot{q}_{F} + \dfrac{\alpha^{2}}{L^{2}}q_{F} \biggl.\biggr] (\beta_{1}-\beta_{2}) = 0
\label{eq:finaleq}
\]
Since we know that $\beta_{1}$ and $\beta_{2}$ are two different fermionic matrices, the only way (\ref{eq:finaleq}) will be equal to zero is when :
\[
\ddot{q}_{F} + \dfrac{\alpha^{2}}{L^{2}}q_{F} = 0
\label{eq:qfeqn}
\]
Further from (\ref{eq:qfeqn}) and (\ref{eq:eleqn1}) we have :
\[
\ddot{q}_{B} + \dfrac{\alpha^{2}}{L^{2}}q_{B} = 0
\label{eq:qbeqn}
\]
Hence, the solutions to the differential equations (\ref{eq:qfeqn}) and (\ref{eq:qbeqn}) which are of the form of simple harmonic oscillator equations with the constants as $\dfrac{\alpha^{2}}{L^{2}}$ are given by :
\[
q_{B} = B_{+} e^{i \, (\alpha\tau/L)} + B_{-} e^{-i \, (\alpha\tau/L)}
\]
\[
q_{F} = F_{+} e^{i \, (\alpha\tau/L)} + F_{-} e^{-i \, (\alpha\tau/L)}
\]
where $B_{+}$ and $B_{-}$ are constant bosonic matrices and $F_{+}$ and $F_{-}$ are constant fermionic matrices respectively.
These four constant matrices between them determine the initial conditions for the sources (mass and charge) and the fields they produce (gravity and Yang-Mills). These four aspects are unified as different aspects of the aikyon. It is interesting that the solution is oscillatory, whereas in the pure gravity case the solution was linear in time evolution.

The action is additive. When there are many aikyons, there is one such action term for every aikyon. An important question arises. Where is the interaction {\it amongst} aikyons? The case that we build in this paper is that the unified theory is geometric, in the same sense that gravity is geometry. Gravity is not an interaction; rather it is a feature of geometry, and motion is geodesic (freefall). In a similar spirit, motion in the unified interaction proposed here is `geodesic', but in a non-commutative geometry. Gauge-fields as 
interaction is only a low energy emergent feature, as we will see below. What actually constitutes as interaction in our theory is entanglement of the aikyons. This is what gives rise to emergent gravity and gauge-fields as distinct forces.

\section{Invariance of the functional form of the Lagrangian under a transformation which leaves the gyromagnetic ratio unchanged}
\label{sec:Action_invariance}
We define the gyromagnetic ratio of an aikyon as the product $\alpha L$. Subsequently we will see that in the emergent dynamics below Planck scale, Planck's constant $\hbar$ emerges. The mass of an aikyon is then defined as $m\equiv \hbar/L c$, giving $L$ the interpretation of Compton wavelength, and showing that our present definition of gyromagnetic ratio coincides with the conventional one  for a fermion satisfying the Dirac equation.
We rewrite below the Lagrangian for an aikyon:
\s[
\mathcal{L} = Tr \biggl[\biggr. \dfrac{L_{p}^{2}}{L^{4}} \biggl\{\biggr. i\alpha \biggl(\biggr. q_{B} + \dfrac{L_{p}^{2}}{L^{2}}\beta_{1} q_{F} \biggl.\biggr) + L \biggl(\biggr. \dot{q}_{B} + \dfrac{L_{p}^{2}}{L^{2}}\beta_{1}\dot{q}_{F} \biggl.\biggr) \biggl.\biggr\}\\
\biggl\{\biggr. i\alpha \biggl(\biggr. q_{B} + \dfrac{L_{p}^{2}}{L^{2}}\beta_{2} q_{F} \biggl.\biggr) + L \biggl(\biggr. \dot{q}_{B} + \dfrac{L_{p}^{2}}{L^{2}}\beta_{2}\dot{q}_{F} \biggl.\biggr) \biggl.\biggr\} \biggl.\biggr]
\s]

This Lagrangian can also be written as :
\s[
\mathcal{L} = Tr \biggl[\biggr. L_{p}^{2} \biggl\{\biggr. i\alpha \biggl(\biggr. \dfrac{q_{B}}{L^{2}} + \dfrac{L_{p}^{2}}{L^{2}}\beta_{1}\dfrac{q_{F}}{L^{2}} \biggl.\biggr) + \biggl(\biggr. \dfrac{\dot{q}_{B}}{L} + \dfrac{L_{p}^{2}}{L^{2}}\beta_{1}\dfrac{\dot{q}_{F}}{L} \biggl.\biggr) \biggl.\biggr\}\\
\biggl\{\biggr. i\alpha \biggl(\biggr. \dfrac{q_{B}}{L^{2}} + \dfrac{L_{p}^{2}}{L^{2}}\beta_{2}\dfrac{q_{F}}{L^{2}} \biggl.\biggr) + \biggl(\biggr. \dfrac{\dot{q}_{B}}{L} + \dfrac{L_{p}^{2}}{L^{2}}\beta_{2}\dfrac{\dot{q}_{F}}{L} \biggl.\biggr) \biggl.\biggr\} \biggl.\biggr]
\s]
We now show that given an aikyon $(q_B, q_F)$ with parameters $(\alpha, L)$, there exists another aikyon $(q_B', q_F')$ with parameters $(\alpha', L')$ whose Lagrangian has the same form as that of the former.
We shall show that under transformations according to $\alpha^{'} L^{'} = \alpha L$ and $L L^{'} = L_{P}^{2}$, the functional form of the Lagrangian remains invariant, which further implies that the functional form of the equations of motion will also remain invariant. The relation $\alpha^{'} L^{'} = \alpha L$ implies that the gyromagnetic ratio remains unchanged. 

The two terms in the brackets that are multiplied in the Lagrangian are the same except for the beta matrices where in the first term, we have $\beta_{1}$ while in the second we have $\beta_{2}$. And since the beta matrices remain invariant under these transformations, we shall only consider the first term and  show the invariance of the functional form of that term under the above mentioned transformations. One can then infer the same about the second term, which implies that the Lagrangian is form invariant under the transformations. Also note that each term inside the Lagrangian, after expanding, is dimensionless.
Under these transformations, the first term of $\mathcal{L}$ becomes,
\[
\mathcal{L}_{first} = Tr \biggl[\biggr. L_{p}^{2} \biggl\{\biggr. \dfrac{i\alpha^{'}L^{'2}}{L_{P}^{2}} \biggl(\biggr. \dfrac{L^{'2}}{L_{P}^{4}}q_{B} + \dfrac{L_{p}^{2}L^{'2}}{L_{P}^{4}}\beta_{1}\dfrac{L^{'2}}{L_{P}^{4}}q_{F} \biggl.\biggr) + \biggl(\biggr. \dfrac{L^{'}}{L_{P}^{2}}\dot{q}_{B} + \dfrac{L_{p}^{2}L^{'2}}{L_{P}^{4}}\beta_{1}\dfrac{L^{'}}{L_{P}^{2}}\dot{q}_{F} \biggl.\biggr) \biggl.\biggr\} \biggl.\biggr]
\]
Now if we try to bring the transformed Lagrangian to the same functional form as the one we started out with, we see that:
\[
\mathcal{L}_{first} = Tr \biggl[\biggr. L_{p}^{2} \biggl\{\biggr. i\alpha^{'} \biggl(\biggr. \dfrac{1}{L^{'2}} \dfrac{L^{'6}}{L_{P}^{6}}q_{B} + \dfrac{L_{p}^{2}}{L^{'2}}\beta_{1}\dfrac{1}{L^{'2}}\dfrac{L^{'10}}{L_{P}^{10}}q_{F} \biggl.\biggr) + \biggl(\biggr. \dfrac{1}{L^{'}}\dfrac{L^{'2}}{L_{P}^{2}}\dot{q}_{B} + \dfrac{L_{p}^{2}}{L^{'2}}\beta_{1}\dfrac{1}{L^{'}}\dfrac{L^{'6}}{L_{P}^{6}}\dot{q}_{F} \biggl.\biggr) \biggl.\biggr\} \biggl.\biggr]
\]
where if we now define $q_{B}^{'} \equiv \dfrac{L^{'6}}{L_{P}^{6}} q_{B}$ and $q_{F}^{'} \equiv \dfrac{L^{'10}}{L_{P}^{10}} q_{F}$, we see that:
\begin{gather}
q_{B}^{'} = \dfrac{L^{'6}}{L_{P}^{6}} q_{B}\\
\implies \dfrac{\partial}{\partial \tau_{n}} q_{B}^{'} = \dfrac{L^{'6}}{L_{P}^{6}} \dfrac{\partial}{\partial \tau_{n}} q_{B}\\
\implies \dfrac{\partial}{\partial \tau_{n}} q_{B}^{'} = \dfrac{L^{'6}}{L_{P}^{6}} \dfrac{\partial \tau}{\partial \tau_{n}} \dfrac{\partial}{\partial \tau} q_{B}
\end{gather}
where $\tau_{n}$ is the new Connes' time, scaled with respect to the earlier $\tau$. And similarly for $q_{F}^{'}$, we have
\begin{gather}
q_{F}^{'} = \dfrac{L^{'10}}{L_{P}^{10}} q_{F}\\
\implies \dfrac{\partial}{\partial \tau_{n}} q_{F}^{'} = \dfrac{L^{'10}}{L_{P}^{10}} \dfrac{\partial}{\partial \tau_{n}} q_{F}\\
\implies \dfrac{\partial}{\partial \tau_{n}} q_{F}^{'} = \dfrac{L^{'10}}{L_{P}^{10}} \dfrac{\partial \tau}{\partial \tau_{n}} \dfrac{\partial}{\partial \tau} q_{F}
\end{gather} 
Now taking $\tau_{n} = \dfrac{L^{'4}}{L_{P}^{4}} \tau$, we get:
\[
\dfrac{\partial \tau}{\partial \tau_{n}} = \dfrac{L_{P}^{4}}{L^{'4}}
\]
which further implies
\[
\overset{n}{\dot{q_{B}^{'}}} = \dfrac{L^{'2}}{L_{P}^{2}} \dot{q}_{B} \hspace{1cm} and \hspace{1cm} \overset{n}{\dot{q_{F}^{'}}} = \dfrac{L^{'6}}{L_{P}^{6}} \dot{q}_{F}
\]
where $\overset{n}{\dot{q_{B}^{'}}}$ denotes $\dfrac{\partial}{\partial \tau_{n}} q_{B}^{'}$ and $\overset{n}{\dot{q_{F}^{'}}}$ denotes $\dfrac{\partial}{\partial \tau_{n}} q_{F}^{'}$.\\

Now writing down the Lagrangian's first term in terms of the new variables, we get
\[
\mathcal{L^{'}}_{first} = Tr \biggl[\biggr. L_{p}^{2} \biggl\{\biggr. i\alpha^{'} \biggl(\biggr. \dfrac{q_{B}^{'}}{L^{'2}} + \dfrac{L_{p}^{2}}{L^{'2}}\beta_{1}\dfrac{q_{F}^{'}}{L^{'2}} \biggl.\biggr) + \biggl(\biggr. \dfrac{\overset{n}{\dot{q_{B}^{'}}}}{L^{'}} + \dfrac{L_{p}^{2}}{L^{'2}}\beta_{1}\dfrac{\overset{n}{\dot{q_{F}^{'}}}}{L^{'}} \biggl.\biggr) \biggl.\biggr\} \biggl.\biggr]
\]
Hence we see that the functional form of the Lagrangian has remained the same under the transformations $\alpha^{'} L^{'} = \alpha L$ and $L L^{'} = L_{P}^{2}$ which leaves the gyromagnetic ratio unchanged. This implies that the equations of motion as well as their solutions, for the first aikyon, can be mapped to those for the second aikyon, using the said transformations, which leave the gyromagnetic ratio unchanged. Later in the paper, we will employ this result to map a strongly quantum system to a strongly classical system. And we explain why the Kerr-Newman black hole has the same gyromagnetic ratio as a Dirac fermion, whereas one would have expected half this value for the black hole, because the black hole is a classical object. This map points to a deep connection between Kerr-Newman black holes and Dirac fermions.

Because it will be relevant later in the paper, we note that the above transformation does scale the momentum by a constant factor. This is because in writing the expression for the action, the time integral changes from being over $\tau$ to being over $\tau'$. This amounts to a scaling of the Lagrangian by the constant $L_P^4 / L^{'4}$ - this of course does not change the equations of motion - but scales the canonical momentum. Thus it is easily shown that $p'_B = (L_P^4 / L^{'4}) p_B$; a result which will be relevant when we use the Dirac equation constructed from $p_B$ to once again establish the invariance of the gyromagnetic ratio.

\section{The generalised  Dirac equation in matrix dynamics}
\label{sec:DE}
In our earlier work, where we proposed the matrix dynamics for gravity coupled to Dirac fermions, we showed that the bosonic and fermionic momenta are constants of motion. This permitted us to write a very useful generalisation of the Dirac equation, relevant for the matrix dynamics. Here we recall those equations, and then in a significant advance, we show how an identical Dirac equation can be set up in the present case as well, when Yang-Mills fields are included, and the implications of this advance.

The matrix dynamics action for a pure gravity aikyon was proposed earlier, and is
\begin{equation}
\frac{S}{C_0}  =  \frac{1}{2} \int \frac{d\tau}{\tau_{Pl}} \; Tr \bigg[\frac{L_P^2}{L^2c^2}\; \left(\dot{q}_B +\beta_1 \frac{L_P^2}{L^2}\dot{q}_F\right)\; \left(\dot{q}_B +\beta_2 \frac{L_P^2}{L^2} \dot{q}_F\right) \bigg]
\end{equation}
It differs from the action proposed in the present paper in that it does not have the Yang-Mills part proportional to $q_B$. 
From the resulting equations of motion it follows that the bosonic and fermionic momenta are constants:
\begin{align}
     p_B  &= \frac{a}{2}\bigg[2\dot{q}_B + \frac{L_P^2}{L^2}(\beta_1 +\beta_2)\dot{q}_F \bigg] = c_1\\ 
    p_F  &= \frac{a}{2} \frac{L_P^2}{L^2}\bigg[\dot{q}_B (\beta_1 +\beta_2)+ \frac{L_P^2}{L^2}\beta_1 \dot{q}_F \beta_2 +  \frac{L_P^2}{L^2}\beta_2 \dot{q}_F \beta_1 \bigg]=c_2
\end{align}
The trace Hamiltonian can be constructed as:
	\begin{eqnarray}
	\textbf{H} = Tr \left[p_{F}\dot{q_{F}}\right] + Tr \left[p_{B}\dot{q_{B}}\right] - Tr \mathcal{L}
	\end{eqnarray}
	For the present case, after plugging in the relevant expressions for momenta, we have for the Hamiltonian:
	\begin{eqnarray}
	\textbf{H}= Tr \Bigl[ \frac{L_{P}^{2}}{L^{4}} \Bigl( \alpha^{2}q_{B}^{2} +\alpha^{2}\frac{L_{P}^{4}}{L^{4}} \beta_{1}q_{F}\beta_{2}q_{F} + {\alpha^{2}} \frac{L_{P}^{2}}{L^{2}} (\beta_{1} + \beta_{2})q_{B}q_{F} \nonumber \\
	+L^{2}\dot{q_{B}}^{2} +\frac{L_{P}^{4}}{L^{2}} \beta_{1}\dot{q_{F}}\beta_{2}\dot{q_{F}} +{L_{P}^{2}}(\beta_{1} + \beta_{2})\dot{q_{B}} \dot{q_{F}}  \Bigr) \Bigr]
	\end{eqnarray}
	In the pure gravity limit, i.e., $\lim \alpha \rightarrow 0$, we recover the form that was originally worked out for the pure gravity case:
	\begin{eqnarray}
		\textbf{H} = Tr \left[\frac{L_{P}^{4}}{L^{2}}\left({\dot{q_{B}}^{2} + \frac{L_{P}^{4}}{L^{4}}\beta_{1}\dot{q_{F}}\beta_{2}\dot{q_{F}} + \frac{L_{P}^{2}}{L^{2}}(\beta_{1} + \beta_{2})\dot{q_{B}}\dot{q_{F}} }\right)\right]
	\end{eqnarray}
Using the same definition for the Dirac operators $D_B$ and $D_F$ as given above, the equation for the constancy of the bosonic momentum operator $p_B$ in the pure gravity case can be written as an eigenvalue equation, as mentioned in the Introduction:
 \begin{equation}
 \left[D_B + D_F\right] \psi =  \frac{1}{L} \bigg(1+ i \frac{L_P^2}{L^2}\bigg)\psi
 \end{equation}
Now in order to come up with a Dirac equation similar to this one, after including Yang-Mills fields,  we define new variables so that the trace Lagrangian looks similar to the one defined above for  gravity.  For this, let us start by proposing the transformation
\[
Q_{B} = e^{i \, (\alpha c\tau/L)} q_{B} \qquad and \qquad Q_{F} = e^{i \, (\alpha c\tau/L)} q_{F}
\]
\[
\implies \dot{Q}_{B} = \dfrac{1}{L} e^{i \, (\alpha c\tau/L)} (i\alpha \, q_{B} + L\, \dot{q}_{B}) \qquad and \qquad \dot{Q}_{F} = \dfrac{1}{L} e^{i \, (\alpha c\tau/L)} (i\alpha \, q_{F} + L\, \dot{q}_{F})
\label{eq:Qdotdefn}
\]
Now by expressing the trace Lagrangian in terms of the new variables defined above, one finds that the Lagrangian becomes :
\s[
\mathcal{L} = Tr \biggl[\biggr. \dfrac{L_{p}^{2}}{L^{4}} \biggl\{\biggr. i\alpha \biggl(\biggr. q_{B} + \dfrac{L_{p}^{2}}{L^{2}}\beta_{1} q_{F} \biggl.\biggr) + L \biggl(\biggr. \dot{q}_{B} + \dfrac{L_{p}^{2}}{L^{2}}\beta_{1}\dot{q}_{F} \biggl.\biggr) \biggl.\biggr\}\\
\biggl\{\biggr. i\alpha \biggl(\biggr. q_{B} + \dfrac{L_{p}^{2}}{L^{2}}\beta_{2} q_{F} \biggl.\biggr) + L \biggl(\biggr. \dot{q}_{B} + \dfrac{L_{p}^{2}}{L^{2}}\beta_{2}\dot{q}_{F} \biggl.\biggr) \biggl.\biggr\} \biggl.\biggr]
\s]
\s[
\implies \mathcal{L} = Tr \biggl[\biggr. \dfrac{L_{p}^{2}}{L^{4}} \biggl\{\biggr. (i\alpha \, q_{B} + L \, \dot{q}_{B}) + \dfrac{L_{p}^{2}}{L^{2}} \beta_{1} (i\alpha \, q_{F} + L \, \dot{q}_{F}) \biggl.\biggr\}\\
\biggl\{\biggr. (i\alpha \, q_{B} + L \, \dot{q}_{B}) + \dfrac{L_{p}^{2}}{L^{2}} \beta_{2} (i\alpha \, q_{F} + L \, \dot{q}_{F}) \biggl.\biggr\} \biggl.\biggr]
\s]

\s[
\implies \mathcal{L} = Tr \biggl[\biggr. \dfrac{L_{p}^{2}}{L^{4}} \biggl\{\biggr. L \, \dot{Q}_{B} \, e^{-i \, (\alpha\tau/L)} + \dfrac{L_{p}^{2}}{L^{2}} \beta_{1} \biggl(\biggr. L \, \dot{Q}_{F} \, e^{-i \, (\alpha\tau/L)} \biggl.\biggr) \biggl.\biggr\}\\
\biggl\{\biggr. L \, \dot{Q}_{B} \, e^{-i \, (\alpha\tau/L)} + \dfrac{L_{p}^{2}}{L^{2}} \beta_{2} \biggl(\biggr. L \, \dot{Q}_{F} \, e^{-i \, (\alpha\tau/L)} \biggl.\biggr) \biggl.\biggr\} \biggl.\biggr]
\s]
Now defining $\dot{\widetilde{Q}}_{B} = \dot{Q}_{B} \, e^{-i \, (\alpha c\tau/L)}$ and $\dot{\widetilde{Q}}_{F} = \dot{Q}_{F} \, e^{-i \, (\alpha c\tau/L)}$, we see that the Lagrangian in terms of $\dot{\widetilde{Q}}_{B}$ and $\dot{\widetilde{Q}}_{F}$ becomes: 
\[
\implies \mathcal{L} = Tr \biggl[\biggr. \dfrac{L_{p}^{2}}{L^{2}} \biggl(\biggr. \dot{\widetilde{Q}}_{B} + \dfrac{L_{p}^{2}}{L^{2}} \beta_{1} \dot{\widetilde{Q}}_{F} \biggl.\biggr) \biggl(\biggr. \dot{\widetilde{Q}}_{B} + \dfrac{L_{p}^{2}}{L^{2}} \beta_{2} \dot{\widetilde{Q}}_{F} \biggl.\biggr) \biggl.\biggr]
\label{eq:tracelag}
\]
We have hence made a transformation to the new variables, from the old ones, in terms of which the trace Lagrangian looks identical in form to the one defined for pure gravity. We believe this is a significant advance, as this new variable $\widetilde{Q}$ represents a unification of gravity and Yang-Mills. Only the constant $L$ appears in the theory now; whereas $\alpha$ is hidden in the phase relating $\widetilde{Q}$ to the earlier variable $q$. In fact it is easy to check that
\begin{equation}
{\dot{\widetilde{Q}}_B} = \frac{1}{L} (i\alpha q_B + L \dot{q}_B); \qquad  {\dot{\widetilde{Q}}_F} = \frac{1}{L} (i\alpha q_F + L \dot{q}_F);
\end{equation}
The gauge potential has been absorbed in the phase, and the new variable behaves like pure gravity; as if to suggest that there is a geometric interpretation for the unified interaction, in the matrix dynamics. This possibly has far-reaching implications. There is a clear analogy with the spectral action principle in non-commutative geometry: there, gauge-fields arise as the so-called inner automorphisms in the larger diffeomorphism group [which includes both gravity and gauge-fields]. The smaller diffeomorphism group describes space-time. For us, this corresponds to the group of global unitary transformations which leaves the action for $\widetilde{Q}$ invariant [as in trace dynamics] and which includes the gauge-interactions as phase-shifts from the original gravity operator $q$. Yet the central difference from their work is that their's is classical physics; whereas we will have quantum theory emergent from our dynamics.

We can  carry on with further analysis and draw similar conclusions for the unified case, as was done for the gravity case.
Now in order to write down the equations for the bosonic momentum $\widetilde{p}_{B}$ and the fermionic momentum $\widetilde{p}_{F}$ for the trace Lagrangian given by (\ref{eq:tracelag}), we need $\dfrac{\partial \mathcal{L}}{\partial \dot{\widetilde{Q}}_{B}}$ and $\dfrac{\partial \mathcal{L}}{\partial \dot{\widetilde{Q}}_{F}}$ the calculations for which are as shown below :
\[
\widetilde{p}_{B} = \dfrac{\partial \mathcal{L}}{\partial \dot{\widetilde{Q}}_{B}} = \dfrac{L_{p}^{2}}{L^{2}} \biggl[\biggr. 2 \dot{\widetilde{Q}}_{B} + \dfrac{L_{p}^{2}}{L^{2}} (\beta_{1} + \beta_{2}) \dot{\widetilde{Q}}_{F} \biggl.\biggr]
\]
\[
\widetilde{p}_{F} = \dfrac{\partial \mathcal{L}}{\partial \dot{\widetilde{Q}}_{F}} = \dfrac{L_{p}^{2}}{L^{2}} \biggl[\biggr. \dfrac{L_{p}^{2}}{L^{2}} \dot{\widetilde{Q}}_{B} \beta_{2} + \dfrac{L_{p}^{2}}{L^{2}} \dot{\widetilde{Q}}_{B} \beta_{1} + \dfrac{L_{p}^{4}}{L^{4}} \beta_{1} \dot{\widetilde{Q}}_{F} \beta_{2} + \dfrac{L_{p}^{4}}{L^{4}} \beta_{2} \dot{\widetilde{Q}}_{F} \beta_{1} \biggl.\biggr]
\]
\[
\implies \widetilde{p}_{F} = \dfrac{\partial \mathcal{L}}{\partial \dot{\widetilde{Q}}_{F}} = \dfrac{L_{p}^{4}}{L^{4}} \biggl[\biggr. \dot{\widetilde{Q}}_{B} (\beta_{1} + \beta_{2}) + \dfrac{L_{p}^{2}}{L^{2}} \beta_{1} \dot{\widetilde{Q}}_{F} \beta_{2} + \dfrac{L_{p}^{2}}{L^{2}} \beta_{2} \dot{\widetilde{Q}}_{F} \beta_{1} \biggl.\biggr]
\]
Here the conjugate momenta, $\widetilde{p}_{B}$ and $\widetilde{p}_{F}$ are constants as the trace Lagrangian is independent of $\widetilde{Q}_{B}$ and $\widetilde{Q}_{F}$ similar to what happened for pure gravity. This implies,
\[
2 \dot{\widetilde{Q}}_{B} + \dfrac{L_{p}^{2}}{L^{2}} (\beta_{1} + \beta_{2}) \dot{\widetilde{Q}}_{F} = C_{1}
\label{eq:dirac1}
\]
\[
\dot{\widetilde{Q}}_{B} (\beta_{1} + \beta_{2}) + \dfrac{L_{p}^{2}}{L^{2}} \beta_{1} \dot{\widetilde{Q}}_{F} \beta_{2} + \dfrac{L_{p}^{2}}{L^{2}} \beta_{2} \dot{\widetilde{Q}}_{F} \beta_{1} = C_{2}
\]
for some $C_{1}$ and $C_{2}$ which are constant bosonic and fermionic matrices respectively. Taking cue from (\ref{eq:dirac1}) in defining the bosonic and fermionic Dirac operators, and recalling the earlier definition of $D_{new}$ we find that 
\[
D_{Bnewi} = \dfrac{1}{L} \dot{\widetilde{Q}}_{B} \qquad and \qquad D_{Fnewi} = \frac{L_P^2}{L^2}\frac{\beta_1+\beta_2}{2Lc} \dot{\widetilde{Q}}_{F}
\label{eq:ddirac}
\]
The operator $D_{Bnewi}$ differs from $D_{Bnew}$ only in that there is an $i$ factor in front of the potential $\dot{q}_{B}$, and an analogous situation exists for $d_{Fnewi}$.  The introduction of the $i$ factor means that the Dirac operator $D_{newi}$ defined from $\widetilde{Q}$ is no longer self-adjoint, although it has the same magnitude as the corresponding Dirac operators made from
self-adjoint $D_{Bnew}$ and $D_{Fnew}$.

Hence we have a constant operator which can be expressed as an eigenvalue equation given by:
\[
\left[D_{Bnewi} + D_{Fnewi}\right] \psi = \lambda \psi
\]
where the eigenvalues $\lambda$ are assumed to be $\mathbb{C}$-numbers [since the operator is bosonic]  and are independent of the Connes' time $\tau$. One also has to bear in mind that although it is deceptive to call a function of $\dot{\widetilde{Q}}_{B}$ and $\dot{\widetilde{Q}}_{F}$ a Dirac operator, it is justified in the sense that this Dirac operator has the same eigenvalues as that defined  on an emergent  space-time manifold in this picture.

Like in the pure gravity case, this can be written as an (important) eigenvalue equation:
\begin{equation}
\left[D_{Bnewi} + D_{Fnewi}\right] \psi =  \frac{1}{L} \bigg(1+ i \frac{L_P^2}{L^2}\bigg)\psi
\end{equation}

\section{Emergence of general relativity and quantum [field] theory as low energy approximations to the Planck scale matrix dynamics}
We have defined the matrix dynamics at the Planck scale, which unifies gravitation and Yang-Mills interactions. It is now necessary to establish that this matrix dynamics can reproduce the low-energy physics - quantum field theory, and classical general relativity. For this we ask, how does does this deterministic dynamics look like, when examined not over Planck time resolution, but after coarse-graining over very many Planck time intervals. In the theory of trace dynamics, this question is answered by employing the conventional techniques of statistical thermodynamics, and we have followed the same approach in our theory. This analysis is described in earlier works \cite{maithresh2019, maithresh2019b}. The outcome is as follows., and falls in two classes: for a low degree of entanglement amongst aikyons, we recover quantum theory, and for high degree of entanglement we recover classical dynamics.

If the length scale $L$ associated with the one or more aikyons in the system is much larger than Planck length, then the anti-self-adjoint part of the system's Hamiltonian can be neglected, and the Adler-Millard charge is anti-self-adjoint.  The emergent dynamics is quantum theory in Connes time. There is no classical space-time yet. Planck's constant emerges as a consequence of equipartition of the Adler-Millard charge, and quantum commutation relations emerge. A bosonic pair $(q_B,p_B)$ satisfies the standard quantum commutation relation, while a  fermionic pair $(q_F, p_F)$ satisfies the quantum anti-commutation relation. Dynamics is described by Heisenberg equations of motion, and equivalently by a Schr\"{o}dinger equation evolving in Connes time. 
In our particular instance dynamics can be described by the pairs $(q_B, p_B)$ and $(q_F, p_F)$ or through the bosonic and fermionic parts of $({\widetilde{Q}}, \widetilde{P}_{\widetilde{Q}})$. The two descriptions are equivalent, and both are a unified description of gravitation and Yang-Mills fields.

The classical limit arises as follows. Suppose a large number $N$ of aikyons get entangled with each other, such that the effective length $L_{eff}\sim L/N$ associated with the entangled system goes below Planck length. In that case, the anti-self-adjoint part of the system's Hamiltonian becomes significant, and the Adler-Millard charge is no longer anti-self-adjoint. Spontaneous localisation results, and a rapid breakdown of superposition of states takes place. The system becomes classical. 
The localisation of the matter part (fermionic) gives rise to the emergence of classical space-time: space-time is operationally defined by the eigenvalues to which the fermions localise.
Using the analysis in our earlier work,  we now demonstrate that the classical limit is general relativity with relativistic material particles and Yang-Mills fields as sources. It seems reasonable now to decouple space-time manifold from gravity: we may think of the manifold as arising from localisation of the fermionic part. The space-time metric is inherent in the non-commutative geometry, in the properties of the Dirac operator. Spontaneous localisation makes the metric manifest in the form that we are familiar with. In the matrix dynamics, gravity and gauge-fields jointly describe the curvature of geometry. In the emergent theory, it appears as if that role is limited only to gravity; however this is only a low energy feature. 

We recall that  in terms of the transformed variable $\widetilde{Q}$ the trace Lagrangian for a single aikyon is of the form:
\[
\mathcal{L} = Tr \biggl[\biggr. \dfrac{L_{p}^{2}}{L^{2}} \biggl(\biggr. \dot{\widetilde{Q}}_{B} + \dfrac{L_{p}^{2}}{L^{2}} \beta_{1} \dot{\widetilde{Q}}_{F} \biggl.\biggr) \biggl(\biggr. \dot{\widetilde{Q}}_{B} + \dfrac{L_{p}^{2}}{L^{2}} \beta_{2} \dot{\widetilde{Q}}_{F} \biggl.\biggr) \biggl.\biggr]
\]
which on expanding gives us four terms which are as follows:
\[
\mathcal{L} = Tr \biggl[\biggr. \dfrac{L_{p}^{2}}{L^{2}} \, \dot{\widetilde{Q}}_{B}^{2} + \dfrac{L_{p}^{4}}{L^{4}} \biggl(\biggr. \dot{\widetilde{Q}}_{B} \beta_{2} \dot{\widetilde{Q}}_{F} + \beta_{1} \dot{\widetilde{Q}}_{F} \dot{\widetilde{Q}}_{B} \biggl.\biggr) + \dfrac{L_{p}^{6}}{L^{6}} \, \beta_{1} \dot{\widetilde{Q}}_{F} \beta_{2} \dot{\widetilde{Q}}_{F} \biggl.\biggr]
\]
We focus on the first term of the trace Lagrangian for now. From (\ref{eq:ddirac}) using the definition of Dirac operators, we see that
\[
Tr\left[{\frac{L_{P}^{2}}{L^{2}}} \, {\dot{{\widetilde{Q}}}_{B}}^{2}\right]=Tr[L_{P}^{2}\, {D^{2}_{Bnewi}}]
\]
where as per our construction, this operator is not self-adjoint and hence does not have real eigenvalues. However, we know 
that the anti-self-adjoint part  causes spontaneous localisation of the self-adjoint part. It is hence natural to assume that under spontaneous localisation, $Tr [D_{Bnewi}^2]$ goes to an eigenvalue of the self-adjoint-operator $D_{Bnew}^2$.
Under spontaneous localisation, each of the aikyons localizes to a specific eigenvalue [different eigenvalues for different aikyons]  reducing the first term of the trace Lagrangian to
\[
Tr[D_{Bnew}^{2}] \rightarrow \lambda_{R}^{2}
\]
If sufficiently many entangled aikyons undergo spontaneous localisation to occupy the various eigenvalues $\lambda_{R}^{i}$ of the Dirac operator $D_{Bnew}$, then we can conclude, from our knowledge of the spectral action in non-commutative geometry \cite{Chams:1997}, that their net contribution to the trace becomes the same as  that of $Tr\; [D_B^2]$ for one aikyon \cite{maithresh2019}
\[
Tr\left[{\frac{L_{P}^{2}}{L^{2}}} \, {\dot{{\widetilde{Q}}}_{B}}^{2}\right] = Tr[L_{P}^{2}\, D_{Bnew}^{2}] = L_{P}^{2} \sum_{i} (\lambda_{R}^{i})^{2} \approx \frac{1}{L_{P}^{2}}\int d^{4}x\sqrt{g}\left({R + L_{P}^{2} \, \alpha^{2} F^{i}_{\mu\nu} F^{\mu\nu i}}\right)
+ {\cal O} (L_P^2)
\]
Hence we see that the eigenvalues of $D_{B}^{2}$ operator sum up to give the combined Yang-Mills and Einstein-Hilbert action terms
\[
Tr\left[{\frac{L_{P}^{2}}{L^{2}}} \, {\dot{{\widetilde{Q}}}_{B}}^{2}\right] = \frac{1}{L_{P}^{2}}\int d^{4}x\sqrt{g}\left({R + L_{P}^{2} \, \alpha^{2} F^{i}_{\mu\nu} F^{\mu\nu i}} \right)
\]
Now let us consider the cross terms in the trace Lagrangian which give us the interaction terms analogous to relativistic charge moving in Yang-Mills and mass-gravity couplings given by:
\[
Tr \left[ \dfrac{L_{p}^{4}}{L^{4}} \biggl(\biggr. \dot{\widetilde{Q}}_{B} \beta_{2} \dot{\widetilde{Q}}_{F} + \beta_{1} \dot{\widetilde{Q}}_{F} \dot{\widetilde{Q}}_{B} \biggl.\biggr) \right] = Tr \left[ \dfrac{L_{p}^{4}}{L^{4}} (\beta_{1} + \beta_{2}) \dot{\widetilde{Q}}_{F} \dot{\widetilde{Q}}_{B} \right]
\]
From (\ref{eq:ddirac}), using the definition of Dirac operators, we see that: 
\[
Tr \left[ \dfrac{L_{p}^{4}}{L^{4}} (\beta_{1} + \beta_{2}) \dot{\widetilde{Q}}_{F} \dot{\widetilde{Q}}_{B} \right] = Tr\left[ \dfrac{L_{p}^{4}}{L^{4}} (\beta_{1} + \beta_{2}) \dfrac{2 L^{4}}{L_{p}^{2}} (\beta_{1} + \beta_{2})^{-1} D_{Fnewi} D_{Bnewi}\right]= Tr\left[ 2{L_{P}^{2}} D_{Fnewi} D_{Bnewi} \right]
\]
Expanding the above expression using the definition of the Dirac operator given by (\ref{eq:ddirac}) and (\ref{eq:Qdotdefn}), we see that the cross terms in the trace Lagrangian  become:
\s[
Tr\left[2 {L_{P}^{2}}D_{F}^{eff}D_{B}^{eff}\right] = Tr \biggl[\biggr. 2 L_{P}^{2} \biggl\{\biggr. \dfrac{L_{p}^{2}}{L^{2}} \biggl(\biggr. \dfrac{\beta_{1}+\beta_{2}}{2L} \biggl.\biggr) \dfrac{1}{L} (i\alpha \, q_{F} + L\, \dot{q}_{F}) \biggl.\biggr\} \biggl\{\biggr. \dfrac{1}{L} \dfrac{1}{L} (i\alpha \, q_{B} + L\, \dot{q}_{B}) \biggl.\biggr\} \biggl.\biggr]\\
= Tr \biggl[\biggr. \dfrac{L_{P}^{4}}{L^{6}} (\beta_{1}+\beta_{2}) \biggl\{\biggr. (i\alpha \, q_{F} + L\, \dot{q}_{F}) (i\alpha \, q_{B} + L\, \dot{q}_{B}) \biggl.\biggr\} \biggl.\biggr]\\
= Tr \biggl[\biggr. \dfrac{L_{P}^{4}}{L^{6}} (\beta_{1}+\beta_{2}) (-\alpha^{2} \, q_{F}q_{B} + i\alpha L \, q_{F} \dot{q}_{B} + i\alpha L \, \dot{q}_{F}q_{B} + L^{2} \, \dot{q}_{F}\dot{q}_{B}) \biggl.\biggr]
\s]
We can ignore the second and thirds terms in the above term of the trace Lagrangian (i.e. the ones with an $i$ factor) because they together form the total time derivative of $q_B q_F$ which hence do not contribute to the variation of the action. Hence the terms we are considering after expanding the cross term of the trace Lagrangian reduce to:
\[
Tr \biggl[\biggr. -\alpha^{2} \dfrac{L_{P}^{4}}{L^{6}} (\beta_{1}+\beta_{2}) q_{F} q_{B} + \dfrac{L_{P}^{4}}{L^{4}} (\beta_{1}+\beta_{2}) \dot{q}_{F}\dot{q}_{B} \biggl.\biggr]
\label{eq:reducedsec}
\]
Proceeding further, we know that the second term in the above equation gives gravity-matter coupling term [as seen in our earlier work \cite{maithresh2019, maithresh2019b}], after spontaneous collapse, giving rise to
\[
Tr[L_{P}^{2} D_{F} D_{B}] = mc \int ds
\]
where $D_{B} = \dfrac{1}{L} \dot{q}_{B}$ and $D_{F} = \dfrac{L_{P}^{2}}{L^{2}} \biggl(\biggr. \dfrac{\beta_{1}+\beta_{2}}{2Lc} \biggl.\biggr) \dot{q}_{F}$. There is one such source term for every aikyon.

This comes about as follows. Spontaneous localisation sends this trace term to $L_p^2 \times {1}/{L_I} \times {1}/{L}$, where $L_I=L^3/L_P^2$. There will be one such term for each STM atom, and analogous to the case of $Tr D_B^2$ we anticipate that the trace over all STM atoms gives rise to the `source term'
\begin{equation}
\hbar\int \sqrt{g} \; d^4x \;\sum_i [ L_p^{-2} \times {1}/{L^i_I} \times {1}/{L^i}]
\end{equation}
Consider the term for one atom. We make the plausible assumption that spontaneous localisation localises the STM atom to a size $L_I$. This is analogous to the resolution length scale.  We know that $L_p^2 L_I = L^3$. We recall that  $L$ is the Compton wavelength $\hbar/ mc$ of the STM atom. Moreover, we propose that the classical approximation consists of replacing the inverse of the spatial volume of the localised particle - $1/L^3$, by the spatial delta function $\delta^3({\bf x} - {\bf x_0})$ so that the contribution to the matter source action becomes
\begin{equation}
\hbar \int \sqrt{g} \; d^4x \; [ L_p^{-2} \times {1}/{L_I} \times {1}/{L}] = mc \int ds
\end{equation}
which of course is the action for a relativistic point particle.

After a few suitable re-definitions, the first term of the cross term in the trace Lagrangian in (\ref{eq:reducedsec}) also correctly gives the Yang-Mills interaction terms as follows. Let
 \begin{eqnarray}
\alpha^{}&=&\frac{q}{\sqrt{\hbar c}}\\
q_{B}&=&A L^{2}/\alpha; \qquad A\longrightarrow A_\mu /\sqrt{\hbar c}\\
q_{F}&=& - L  (\beta_{1}+\beta_{2})^{-1} \frac{dx^{\mu}}{ds}
\end{eqnarray}
So that the first term after the spontaneous localization (heat kernel expansion using $L_{P}^{2}$ as a parameter) gives us :
\[
Tr \biggl[\biggr. L_{P}^{2}{\biggl(\biggr. {-\alpha^{2} \dfrac{L_{P}^{2}}{L^{6}}}(\beta_{1} + \beta_{2}) q_{F}q_{B}\biggl.\biggr)}\biggl.\biggr] = \frac{q}{c}\int ds A_{\mu}\frac{dx^{\mu}}{ds} = \frac{q}{c}\int dx_{\mu}A^{\mu}
\]
So we have been able to make progress towards a unified description of Yang-Mills and gravity by recovering all the relevant terms. Hence we have from the trace Lagrangian after spontaneous localization
\s[
\sum Tr \biggl[\biggr. \dfrac{L_{p}^{2}}{L^{2}} \, \dot{\widetilde{Q}}_{B}^{2} + \dfrac{L_{p}^{4}}{L^{4}} \biggl(\biggr. \dot{\widetilde{Q}}_{B} \beta_{2} \dot{\widetilde{Q}}_{F} + \beta_{1} \dot{\widetilde{Q}}_{F} \dot{\widetilde{Q}}_{B} \biggl.\biggr) + \dfrac{L_{p}^{6}}{L^{6}} \, \beta_{1} \dot{\widetilde{Q}}_{F} \beta_{2} \dot{\widetilde{Q}}_{F} \biggl.\biggr]\\
= \frac{1}{L_{P}^{2}}\int d^{4}x\sqrt{g}\left({R + L_{P}^{2} \, \alpha^{2} F^{i}_{\mu\nu} F^{\mu\nu i}}\right) + \sum \frac{q}{c}\int dx_{\mu}A^{\mu} +\sum  mc \int ds
\s]
The sum denotes sum over all aikyons. As expected, in the emergent theory, the sources are summed. But the emergent classical  fields [gravity and Yang-Mills] are a result of the net contribution of all aikyons.

The additional terms coming from the trace action, at higher order [order $L_{P}^2$] are:
\[\int d^4x\; \sqrt{g}\; 
\frac{L_P^2}{L^8}\  \beta_1  (i\alpha q_F + L \dot{q}_F)\times \beta_2 (i\alpha q_F + L \dot{q}_F) \\
= \int d^4x\; \sqrt{g}\; \frac{L_P^2}{L^6}\  \left[\dot{q}_F^2 - \frac{\alpha^2}{L^2} q_F^2\right]
\]
(two of the four  terms cancel out); there being one such contribution for each aikyon.

Symbolically, spontaneous localisation in the matrix dynamics sends the total action of the aikyons to:
\begin{equation}
S = \int d\tau \sum_i Tr D^2_i   {\mathbf \rightarrow}  \int d\tau  \int d^4x\; \sqrt{g} \; \bigg [\frac{c^3}{2G}R + { \, \alpha^{2} F^{i}_{\mu\nu} F^{\mu\nu i}}+\; \sum_i  \delta^3({\bf x} - {\bf x_0(s)}) \left( cm_i+ \frac{q_i}{c} A_{\mu}u^\mu\right) \bigg]
\end{equation}
This generalises the  earlier result of the pure gravity case shown above in Eqn.(\ref{pg}). One could also ask, just as the source for gravity is the mass $m$, and the source for the gauge-field is the charge $q$, then in the original matrix dynamics, what is the `source' for the unified dynamical variable $\widetilde{Q}_{B}$? An inspection of this equation above suggests that the source is $(1/L+\alpha^2 q_B q_F/L^2)$. This has the nature of a charge-induced correction to mass, and the implications of this observation  remain to be investigated. Also intriguing is the role of the fifth time-like dimension $\tau$ which stays in the background.

We can now explain why the Kerr-Newman black hole has the same gyromagnetic ratio as a Dirac fermion \cite{Adamo}. Let us write the modified Dirac equation for an aikyon $(q_B,q_F)$ with parameters $(L,\alpha)$. And another identical equation for a different aikyon $(q_B', q_F')$ with parameters $(\alpha', L')$ such that $L'=L_P^2 /L$ and $\alpha' L' = \alpha L$:
\begin{equation}
\left[D_{Bnewi} + {D}_{Fnewi}\right] \psi =  
\frac{1}{L} \bigg(1+ i \frac{L_P^2}{L^2}\bigg)\psi
\end{equation}
\begin{equation}
\left[D'_{Bnewi} + { D}'_{Fnewi}\right] \psi =  
= \frac{1}{L'} \bigg(1+ i \frac{L_P^2}{L^{'2}}\bigg)\psi
\end{equation}
Written explicitly, these equations become
\begin{equation}
\left[ \dfrac{1}{L} \dot{\widetilde{Q}}_{B} + \frac{L_P^2}{L^2}\frac{\beta_1+\beta_2}{2Lc} \dot{\widetilde{Q}}_{F}
\right] \psi =  
\frac{1}{L} \bigg(1+ i \frac{L_P^2}{L^2}\bigg)\psi
\end{equation}
\begin{equation}
\left[ \dfrac{1}{L'} \dot{\widetilde{Q}}'_{B} + \frac{L_P^2}{L^{;2}}\frac{\beta_1+\beta_2}{2L'c} \dot{\widetilde{Q}}'_{F}
\right] \psi =  
\frac{1}{L'} \bigg(1+ i \frac{L_P^2}{L^{'2}}\bigg)\psi
\end{equation}
and hence that
\begin{equation}
\left[ \dfrac{1}{L}  \frac{1}{L} \left(i\alpha q_B + L \frac{d{q}_B}{d\tau}\right) + \frac{L_P^2}{L^2}\frac{\beta_1+\beta_2}{2Lc}  \frac{1}{L} \left(i\alpha q_F + L \frac{{q}_F}{d\tau}\right)
\right] \psi =  
\frac{1}{L} \bigg(1+ i \frac{L_P^2}{L^2}\bigg)\psi
\end{equation}
\begin{equation}
\left[ \dfrac{1}{L'}  \frac{1}{L'} \left(i\alpha' q'_B + L' \frac{d{q}'_B}{d\tau'}\right) + \frac{L_P^2}{L^{'2}}\frac{\beta_1+\beta_2}{2L'c}  \frac{1}{L'} \left(i\alpha' q'_F + L' \frac{{q}'_F}{d\tau'}\right)
\right] \psi =  
\frac{1}{L'} \bigg(1+ i \frac{L_P^2}{L^{'2}}\bigg)\psi
\end{equation}

  It is then  shown, as was done for the Lagrangian,  that the transformation
\begin{equation}
q_{B}^{'} \equiv \dfrac{L^{'6}}{L_{P}^{6}} q_{B}; \qquad  q_{F}^{'} \equiv \dfrac{L^{'10}}{L_{P}^{10}} q_{F}; \qquad \tau' = 
\dfrac{L^{'4}}{L_{P}^{4}} \tau
\end{equation}
maps the first of these Dirac equations to the second one; with one anticipated difference. An extra factor of $L_P^4/L^{'4}$ multiplies the eigenvalue on the right hand side of the second equation, after the transformation. This, as we noted earlier, happens because of the scaling of the constant bosonic momentum because of the scaling from $\tau$ to $\tau'$.   Now, if we assume that $L$ is much larger than Planck length, then the imaginary part of the eigenvalue in the first equation is negligible, and the aikyon is quantum in nature, and satisfies the same Dirac equation as a Dirac fermion. Also, $L'$ as defined above is much smaller than Planck length. Hence the second aikyon undergoes spontaneous localisation and is classical in nature. Its dynamics is hence described by the classical Einstein-equations coupled to relativistic point particles and Yang-Mills fields. If the spontaneously localised object possesses a non-zero electric charge $e_{KN}$, then this classical solution is a Kerr-Newman black hole. For it, $\alpha ' L'$ is $(e_{KN}/m_{KN}) (\hbar/c^3/2)$, because mass for the aikyon has been defined through the relation $L=\hbar /mc$. Thus the product $\alpha ' L'$ is proportional to the gyromagnetic ratio $e_{KN}/m_{KN}$ of the Kerr-Newman black hole. Similarly, in the first Dirac equation for a Dirac fermion, $\alpha L$ is proportional to the fermion's gyromagnetic ratio $e/m$ with the same proportionality constant. Hence it follows that a Kerr-Newman black hole can be mapped by a transformation, to a Dirac fermion, such that the transformation leaves the gyromagnetic ratio unchanged. The black hole is dual to a Dirac fermion. This helps understand why  a black hole has parameters analogous to those of an elementary particle, despite the former being classical, and the latter being quantum,

Given the classical space-time background produced by the localised fermions, one can arrive at quantum field theory for the unlocalised degrees of freedom, just as is done in the theory of trace dynamics. The Lagrangian for these `quantum' degrees of freedom is already known, and if we neglect the gravity aspect of the aikyons, we have a Lagrangian for their fermionic and Yang-Mills aspect. This can be expressed as the standard Lagrangian for Dirac fermions and gauge-fields. The Connes time evolution can be exchanged with time evolution in the time of ordinary space-time. In our opinion though, a better approach is to relate to quantum field theory in the Horwitz-Stueckelberg formalism \cite{stueck}, which is manifestly Lorentz covariant. In that case, one can identify Connes time with the absolute time parameter of the Stueckelberg formulation of quantum field theory.

\section{Discussion}
We have used our newly proposed matrix dynamics to explain the counter-intuitive fact that  a Kerr-Newman black hole, despite being classical, has the same gyromagnetic ratio as a Dirac fermion, both being twice the classical value. This has been achieved by showing that a solution of the Dirac equation describing a fermion can be mapped to a solution of Einstein equations describing a charged rotating black hole having the same gyromagnetic ratio as the fermion. We believe this result is support for the validity of this matrix dynamics. Earlier we have used this matrix dynamics to derive the Bekenstein-Hawking entropy for a Schwarzschild black hole, from the microstates of its constituent aikyons \cite{maithresh2019b}. We have also proposed that dark energy is a large scale quantum gravitational phenomenon \cite{Singh:DE}, which is due to an enormous collection of ultra-light aikyons which have not undergone spontaneous localisation. We have also predicted the holographic Karolyhazy uncertainty relation, as a consequence of our matrix dynamics \cite{Singh:KL}.

Through the introduction of the variable $\widetilde{Q}$ we have provided a unification of gravity and gauge-fields, as well as their source charges $L$ and $\alpha$. These together become the bosonic (gravity + Yang-Mills) and fermionic (sources) parts of the aikyon. Moreover, the dynamics of the aikyon described by $\widetilde{Q}$ is that of a free particle. In that sense the aikyon obeys the equivalence principle, while evolving in Connes time at the Planck scale. Thus we have a (non-commutative) geometric description of the unified interaction. After spontaneous localisation, the unification is lost. It remains to be seen if the standard model of particle physics is a consequence of this unified framework.

Another important test of our matrix dynamics is that it must provide an understanding for the origin of spin. The action for the aikyon provides all that there is to know. The concept of mass emerges at low energies, defined in terms of Planck's constant and the fundamental length $L$, and without reference to space-time.  Thus there seems to be no reason why a definition of spin should not emerge too, perhaps without reference to space-time or Lorentz invariance. We do have the operators $(q, p)$ in the matrix dynamics which describe `position' and `momentum'. Is it possible to define spin from them? Does one have to introduce torsion in the definition of the aikyon? We need to show that if the d.o.f. is fermionic and obeys an anti-commutation relation, the associated spin is half-integral. And if it is bosonic and obeys a commutation relation, the associated spin is integral. This issue is under investigation. 

Further details of this matrix dynamics, also known as Spontaneous Quantum Gravity, are available by way of reviews, in \cite{Singh:sqg,Singh:2019,Singh2019qf}. The theory of spontaneous localisation has been reviewed in \cite{RMP:2012, Bassi:03}. 
The problem of time in quantum theory is discussed in \cite{Singh:2012} and also in \cite{Singh:2017} and \cite{QTST:2017}.

\bigskip
\noindent {\bf Acknowledgements:} MSM and AP would like to thank the Tata Institute of Fundamental Research, Mumbai - where this research was carried out - for its kind hospitality.

\newpage

\centerline{\bf REFERENCES}
\bibliographystyle{unsrt}
\bibliography{biblioqmtstorsion}

\end{document}